\documentclass[11pt,a4paper]{iopart}
\usepackage[utf8]{inputenc}
\expandafter\let\csname equation*\endcsname\relax

\expandafter\let\csname endequation*\endcsname\relax 
\usepackage{amsmath}
\usepackage{amsfonts}
\usepackage{amssymb}
\usepackage{amsthm}
\usepackage{newlfont}
\usepackage{mathrsfs}
\usepackage{subfigure}
\usepackage{graphicx}
\usepackage{todonotes}
\usepackage{hyperref}

\newcommand{\opunit}{\text{1}\kern-0.22em\text{l}}

\usepackage{soul}

\newcommand{\id}{\textrm{d}}

\def\bea{\begin{eqnarray}}
\def\eea{\end{eqnarray}}

\def\ba{\begin{array}}
\def\ea{\end{array}}

\def\la{\langle}
\def\ra{\rangle}

\usepackage{etoolbox}
\makeatletter
\newrobustcmd{\fixappendix}{%
  \patchcmd{\l@section}{1.5em}{7em}{}{}%
  \patchcmd{\l@subsection}{2.3em}{7em}{}{}%
}
\makeatother

\begin{document}
 
%\title{Inertial Run-and-Tumble particles in one-dimension}

\title{Dichotomous acceleration process in one dimension: Position fluctuations}
\author{Ion Santra}
\address{Raman Research Institute, Bengaluru 560080, India}
\author{ Durgesh Ajgaonkar}
\address{Indian Institute of Science Education and Research, Pune 411008, India}
\author{Urna Basu} 
\address{S. N. Bose National Centre for Basic Sciences, Kolkata 700106, India}

\begin{abstract}
 We study the motion of a one-dimensional particle which reverses its direction of acceleration stochastically. We focus on two contrasting scenarios, where the waiting-times between two consecutive acceleration reversals are drawn from (i) an exponential distribution and (ii) a power-law distribution  $\rho(\tau)\sim \tau^{-(1+\alpha)}$. We compute the mean, variance and short-time distribution of the position $x(t)$ using a trajectory-based approach. We show that, while for the exponential waiting-time, $\la x^2(t)\ra\sim t^3$ at long times, for the power-law case, a non-trivial algebraic growth $\la x^2(t)\ra \sim t^{2\phi(\alpha)}$ emerges, where $\phi(\alpha)=2$, $(5-\alpha)/2,$ and $3/2$ for $\alpha<1,~1<\alpha\leq 2$ and $\alpha>2$, respectively. 
 Interestingly, we find that the long-time position distribution in case (ii) is a function of the scaled variable $x/t^{\phi(\alpha)}$ with an $\alpha$-dependent scaling function, which has qualitatively very different shapes for $\alpha<1$ and $\alpha>1$. In contrast, for case (i), the typical long-time fluctuations of position are Gaussian.
 
 \end{abstract}

\noindent\rule{\hsize}{2pt}
\tableofcontents
\noindent\rule{\hsize}{2pt}
\title[Dichotomous acceleration process]
\maketitle

\section{Introduction}
Random acceleration process (RAP) refers to the motion of an undamped particle subjected to a random force, modeled by a Gaussian white noise~\cite{wanguhlenbeck}. Such processes arise in various physical situations, including the short-time regime of different kinds of active motions~\cite{Basu2018,Santra2021}, granular collapse~\cite{Cornell1998,Swift1999}, particle motion in a sheared medium~\cite{Dean2021}. It also appears in the context of fluctuating Gaussian interface~\cite{Majumdar2001}, equilibrium dynamics of semi-flexible polymers~\cite{Burkhardt1993} as well as Burgers equation with Brownian initial velocity~\cite{Valageas2009}.

In one dimension the position $x(t)$ of such a randomly accelerated particle follows the Langevin equation,
\begin{align}
\ddot{x}(t)=\eta(t),\label{eq:rap1}
\end{align}
where $\eta(t)$ is a Gaussian white noise with $\la \eta(t) \ra=0$ and $\la\eta(t)\eta(t')\ra=\delta(t-t')$.
 The quantity $x(t)$ also measures the area under a Brownian curve, which is a well-studied problem in mathematics literature~\cite{Goldman1971,Sinai1992,Lachal1994}. Despite the Gaussian nature of the process, it shows an anomalous persistence behavior, as well as a host of related non-trivial phenomena, which have been studied extensively~\cite{McKean1962,Burkhardt1993,
 Burkhardt1997,Burkhardt2000,Smedt2001,
 Burkhardt2007,Majumdar2010,Burkhardt2014,Hilhorst2008,Reymbaut2011,Boutcheng2016,Singh2020}.
 
A natural generalization of the RAP is the case where the random force has a finite memory, eg., where the force is a coloured noise instead of a white noise. A straightforward way to dynamically generate such a scenario is to consider a dichotomous acceleration process (DAP),
\bea
\ddot x (t)= a_0 \sigma(t), \label{eq:model}
\eea  
where $\sigma(t)$ is a dichotomous noise which flips stochastically between $1$ and $-1$;  the waiting time $\tau$ between two consecutive flips are drawn from a distribution $\rho(\tau)$.
 The simplest scenario is when the waiting time distribution is exponential, i.e., $\rho(\tau)=\beta e^{-\beta \tau}$. This corresponds to a Markovian flipping protocol where the flips happen at a constant rate $\beta$. The dichotomous noise with a Markovian flipping protocol appears in Telegraph processes~\cite{Weiss2002,balaBook}, persistent random walks~\cite{Othmer1988,Masoliver1992} and more recently in context of run-and-tumble motions~\cite{Tailleur2008,Malakar2017}. The inverse of the flipping rate $\beta$ sets the characteristic time-scale of the force as is evident from the auto-correlation $\la\sigma(t)\sigma(t')\ra=e^{-2\beta|t-t'|}$. The above dynamics can be thought of as a simple active velocity process~\cite{Razin2019,Banerjee2020,Smith2022}, where tumbling leads to the reversal of acceleration. In fact, \eref{eq:model} with exponential waiting time distribution is a special case of the generalized run-and-tumble process introduced by Dean et. al. in~\cite{Dean2021}, where they have characterized the position distribution in the long-time regime from an exact computation of the generating function. However, the short-time position fluctuations of the DAP even with the Markovian flipping protocol is still unexplored.

In a more general sense, one can consider the DAP with a non-Markovian flipping protocol in which case the waiting-time distribution is not exponential. We focus on the position fluctuations and compute the first two moments using a trajectory based approach. We illustrate this framework for a specific non-Markovian flipping protocol, namely, when the waiting-time is drawn from a power-law distribution $\rho(\tau)=\alpha\tau_0^{\alpha}/\tau^{1+\alpha}$ for $\tau\in[\tau_0,\infty)$, known as the Pareto distribution~\cite{Arnold2008}. In this case, we show that, at late times, the average position grows algebraically as $\la x (t)\ra\sim t^{\theta(\alpha)}$, where $\theta(\alpha)=2-\alpha$ for $\alpha<1$ and $1$ for $\alpha>1$. We also find that $\alpha=1$ is a special case where a logarithmic correction appears and $\la x (t)\ra\sim t\log t$. The second moment also grows algebraically---$\la x^2(t) \ra\sim t^{2\phi(\alpha)}$, where $\phi(\alpha)=2$, $(5-\alpha)/2,$ and $3/2$ for $\alpha<1,~1<\alpha\leq 2$ and $\alpha>2$ respectively.

The trajectory based approach also provides a way to characterize the short-time position distribution as a perturbative series in number of flips. We find that when $\rho(\tau)$ is exponential, the position  probability density increases monotonically away from the origin (initial position), diverging at $x=\pm a_0t^2/2$. On the other hand, for power-law distributed waiting-times, additional peaks appear near the origin, which can also be understood using the same approach. 

At late-times, the position distribution shows very different behavior for exponential and power-law waiting-times. In the former case, at $t\gg \beta^{-1}$, the typical position fluctuations of DAP are similar to that of an RAP, giving rise to a Gaussian distribution with width $\sim t^{3/2}$. A distinctly different behavior is seen for the late-time ($t\gg\tau_0$) position distribution in the presence of a power-law distributed waiting-time. We show, using  numerical simulations, that, in this regime, the position distribution has a scaling form,
\begin{align}
P(x,t)=t^{-\phi(\alpha)}{\cal F}_\alpha\Bigg( \frac{x}{t^{\phi(\alpha)}}\Bigg).
\label{plaw_scaling}
\end{align} 
The scaling function ${\cal F}_\alpha(z)$ has qualitatively different shapes for $\alpha<1$ (with a minimum at $z=0$ and divergences at $z=\pm 1/2$), $1<\alpha<2$ (single but non-Gaussian peak at $z=0$) and $\alpha>2$. For $\alpha>2$, the scaling function evenually approaches a Gaussian, which can be heuristically understood using a central limit theorem-like argument.

The paper is organized as follows. In the next section we discuss the trajectory based approach for a DAP with arbitrary waiting time distribution. In Sections~\ref{sec:exp} and ~\ref{sec:plaw} we discuss the behavior of the position fluctuations in the short-time and long-time regimes for the exponential and power-law waiting time distributions, respectively. We conclude with some open questions in Section~\ref{sec:conc}.
\begin{figure}
\centering\includegraphics[width=0.55\hsize]{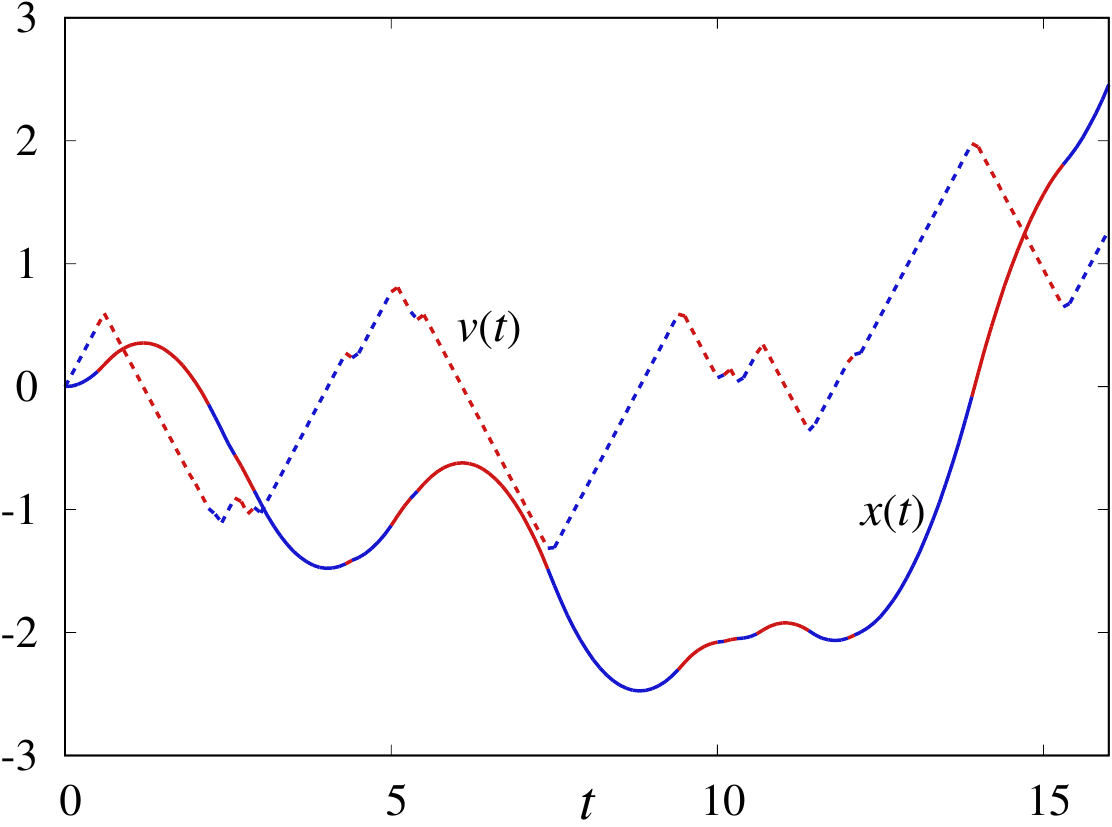} 
\caption{Typical trajectory of a DAP with a constant flip rate $\beta=2$ and $a_0=1$. The blue and red segments denote $\sigma(t)=1$ and $\sigma(t)=-1$, respectively.}\label{fig:traj}
\end{figure}
%%%%%%%%%%%%%%%%%%%%%%%%%%%%%%%%%%%%%%%%%%%

\section{Trajectory based approach}\label{sec:formalism}

In this section we discuss a trajectory based approach to characterize the position fluctuations of the DAP. To this end it is convenient to rewrite 
Eq.~\eqref{eq:model} as,
\bea
\dot x(t) &=& v(t), \cr
\dot v(t) &=& a_0 \sigma(t), \label{eq:xv}
\eea 
where $v(t)$ denotes the instantaneous velocity of the particle. Clearly, $v(t)$ undergoes a L\'evy walk with a arbitrary waiting time distribution~\cite{VasilyRev,Das2021}. In fact, \eref{eq:xv} can be thought of as an undamped generalized inertial L\'evy process~\cite{Lu2011}, or describing the time evolution of the area under the usual overamped L\'evy process. Figure~\ref{fig:traj} shows the typical time evolutions of $x(t)$ and $v(t)$ for the exponential $\rho(\tau)$, i.e., the Markovian case---between two consecutive  flips of $\sigma(t)$ the velocity changes linearly with time, and the position changes quadratically. For different waiting-time distributions the trajectories look qualitatively similar, although, obviously, the typical length of the quadratic pieces changes.

\begin{figure}
\centering \includegraphics[width=12 cm]{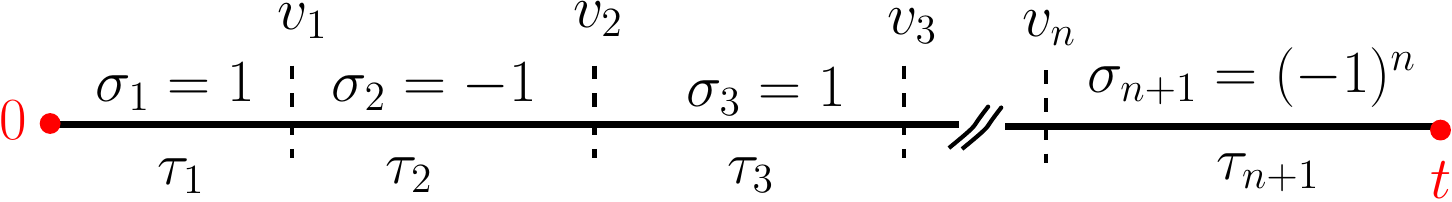}
\caption{Schematic representation of a trajectory with $n$ flips of the acceleration.}\label{fig:schem}
\end{figure}

Let us consider a trajectory of the particle during the interval $[0,t]$ with $n$ flips of $\sigma.$ Let $\tau_i$ denote the interval between
$i^{th}$ and $(i-1)^{th}$ flips; clearly, $\sum_{i=1}^{n+1}\tau_i =t$. Moreover, let $\sigma_i$ denote the constant force during this interval [see \fref{fig:schem} for a schematic representation]. Then, the position increment during this interval is given by,
\bea
\Delta x_i = v_{i-1} \tau_i + \frac {a_0}2 \sigma_i \tau_i^2,
\eea
where $v_{i-1}$ indicates the velocity {\it at the end} of the $(i-1)^{th}$ interval. From Eq.~\eqref{eq:xv}, obviously, 
\bea
v_i = a_0 \sum_{j=1}^i \sigma_j \tau_j.
\eea
In the following we assume that the particle starts from the origin initially. The final position $x(t)$ after time $t$ is then obtained by summing over the $n+1$ increments. 

Combining the above equations we get the final position of the particle at time $t,$
\bea
x &= & \sum_{i=1}^{n+1} \Delta x_i 
= a_0\sum_{i=2}^{n+1} \sum_{j=1}^{i-1} \sigma _j \tau_j \tau_i + \frac {a_0}2 \sum_{i=1}^{n+1} \sigma_i \tau_i^2 ,
\eea
with $\sigma_i = (-1)^{i-1} \sigma_1.$ It is convenient to recast this expression as a symmetric bilinear form,
\begin{align}
x=\frac {a_0}2 \sum_{i,j=1}^{n+1} \chi_{ij} \tau_i \tau_j, \quad  \text{with} \quad \chi_{ij} = \sigma_i \delta_{ij} +(1-\delta_{ij}) \sigma_{\min(i,j)}.
\label{eq:defn_xchi}
\end{align}
Note that, the displacement $x$ after time $t$ is strictly bounded in the region $-x_0 \le x \le x_0$, where $x_0= a_0t^2/ 2 $, and hence the position distribution $P(x,t)$ is supported only in the region  $x\in[-x_0,x_0]$.

Let us consider the general case where $\sigma$ flips follow an arbitrary waiting time distribution $\rho(t)$ --- the probability that $\sigma$ has not flipped for a duration $t$ is then given by $f(t) = \int_t ^ \infty d \tau \rho(\tau).$ The probability density corresponding to a given $\sigma$ trajectory $\{\sigma_i, \tau_i; i=1,2, \cdots n+1\}$ is then,
\begin{align}
{\cal P}(\{\sigma_i, \tau_i\}) = f(\tau_{n+1}) \prod_{i=1}^n \rho(\tau_i).  
\end{align}
The position distribution, in turn, can formally be expressed as
%\bea
%\fl P(x,t) = \sum_{\sigma_1 = \pm 1} \mu(\sigma_1) \sum_{n=0}^\infty \int \prod_{i=1}^{n}\id \tau_i \id \tau_{n+1}\rho(\tau_i) f(\tau_{n+1}) \delta(t - \sum_{i=1}^{n+1} \tau_i) \delta\left(x-  \frac {a_0}2 \sum_{i,j=1}^{n+1} \chi_{ij} \tau_i \tau_j\right)~~\label{eq:Pxt_gen}
%\eea
%
\begin{align}
P(x,t) = \sum_{\sigma_1 = \pm 1} \mu(\sigma_1) \sum_{n=0}^\infty P_{n,\sigma_1}(x,t), ~~\label{eq:Pxt_gen}
\end{align}
where $\mu(\sigma_1)$ denotes the probability that the initial direction of the force is $\sigma_1$, and
\begin{align}
P_{n,\sigma_1}(x,t)=\int \prod_{i=1}^{n}\id \tau_i \id \tau_{n+1}\rho(\tau_i) f(\tau_{n+1}) \delta(t - \sum_{i=1}^{n+1} \tau_i) \delta\left(x-  \frac {a_0}2 \sum_{i,j=1}^{n+1} \chi_{ij} \tau_i \tau_j\right),
\label{eq:pnsigma_gen}
\end{align}
denotes the probability that the final position is $x$ after exactly $n$ number of flips starting with initial acceleration sign $\sigma_1$. It is clear from \eref{eq:defn_xchi}, that $\chi_{ij}\to-\chi_{ij}$ for $\sigma_1\to-\sigma_1$. This implies that,
\begin{align}
P_{n,-\sigma_1}(x,t)=P_{n,\sigma}(-x,t).
\label{pnsymmetry}
\end{align}
This symmetry can be used to obtain the full distribution by computing $P_{n,+}(x,t)$ only. 
In the following, we first compute the mean and the variance of the DAP using Eqs.~\eqref{eq:Pxt_gen} and \eqref{eq:pnsigma_gen}.

%\end{widetext}

%However, this formal expression is not very useful directly as the $\delta$-functions make it difficult to allow ...

\subsection{Moments of position distribution}

The position moments can be formally obtained from Eq.~\eqref{eq:Pxt_gen},
\begin{align} 
 \la x^k(t)& \ra = \int dx\, x^k P(x,t) \\
&= \Big(\frac{a_0}{2}\Big)^k \sum_{\sigma_1 = \pm 1} \mu(\sigma_1)\sum_{n=0}^\infty \int \prod_{i=1}^{n}\id \tau_i  \id \tau_{n+1}\rho(\tau_i) f(\tau_{n+1}) \delta\left(t - \sum_{i=1}^{n+1} \tau_i\right) \left(\sum_{i,j=1}^{n+1} \chi_{ij} \tau_i \tau_j \right)^k.\nonumber
\end{align}
It is convenient to take the Laplace transform of the above expression with respect to time $t$,
\begin{align}
m_k(s) &= \int_0^t dt~ e^{-st} \la x^k(t) \ra \cr
&=\Big(\frac{a_0}{2}\Big)^k \sum_{\sigma_1 = \pm 1} \mu(\sigma_1)\sum_{n=0}^\infty \int_0^\infty \prod_{i=1}^{n}\id \tau_i \id \tau_{n+1} \rho(\tau_i) f(\tau_{n+1}) e^{-s\sum_{i=1}^{n+1} \tau_i} \left(\sum_{i,j=1}^{n+1} \chi_{ij}\tau_i \tau_j \right)^k. \cr \label{eq:mks}
\end{align}
In principle, the Laplace transform of any moment can be computed using Eq.~\eqref{eq:mks} and here we show the explicit computations of the first two moments. For $k=1$, the Laplace transform of the average position $\la x(t) \ra$, from Eq.~\eqref{eq:mks}, is given by,
\begin{align}
m_1(s) = \frac{a_0}2 \sum_{\sigma_1=\pm 1} \mu(\sigma_1) \sum_{n=0}^{\infty} \int_0^\infty \prod_{i=1}^{n} d\tau_i d \tau_{n+1} \rho(\tau_i) f(\tau_{n+1}) e^{-s\sum_{i=1}^{n+1} \tau_i} \sum_{i,j=1}^{n+1}\chi_{ij} \tau_i \tau_j, 
\end{align}
where the symmetric matrix $\chi_{ij}$ is defined in Eq.~\eqref{eq:defn_xchi}. The sum can be evaluated using the explicit form of $\chi_{ij}$ and we get [see \ref{app:moments} for details],
\bea 
m_1(s) = a_0 \la \sigma_1 \ra \frac {1- g(s)}{s^3 (1+g(s))}, \label{eq:xav_laplace}
\eea 
where $g(s)$ is  the Laplace transform of the waiting time  distribution,
\bea 
g(s) &=& \int_0^\infty dt~ e^{-st}\rho(t), \label{eq:g-def}
%h(s) &=& \int_0^\infty dt~ e^{-st} f(t) = \frac 1s \big[ 1- g(s) \big] \label{eq:hs}
\eea
and $a_0\la\sigma_1\ra=a_0\sum_{\sigma_1=\pm1}\sigma_1\,\mu(\sigma_1)$ denotes the average initial acceleration.
Proceeding similarly, the Laplace transform of the second moment can also be computed [see \ref{app:moments} for details],
\bea
 m_2(s) = 2a_0^2 \left[\frac 3{s^5} + \frac{s g''(s) - 6 g'(s)}{s^4(g^2(s)-1)} - \frac{2 g'^2(s)}{s^3(g^2(s)-1)(g(s)+1)}\right]. \label{eq:x2av_laplace}
\eea
To proceed further, we need to specify the waiting time distribution. In the following sections we consider two particularly interesting scenarios, namely, exponential and power-law waiting time distributions.

%\begin{widetext}
%
%\bea 
%m_2(s) = 2a_0^2 \frac{3(g^3+g^2-1)+s(sg''-2g'(sg'+3) )+g(s^2g''-6sg'-3)}{s^5(g-1)(g+1)^2}
%\eea 
%
%\end{widetext} 

%\begin{widetext}
%\bea 
%\la x^k(t) \ra &=& \int dx\, x^k \sum_{\sigma_1 = \pm 1} \mu(\sigma_1)\sum_{n=0}^\infty \int \prod_{i=1}^n\id \tau_i \id \tau_{n+1} ~\rho(\tau_i) f(\tau_{n+1}) \delta(t - \sum_{i=1}^{n+1} \tau_i) \delta\left(x- a_0\sum_{i=2}^{n+1} \sum_{j=1}^{i-1} \sigma _j \tau_j \tau_i - \frac {a_0}2 \sum_{i=1}^{n+1} \sigma_i \tau_i^2\right) 
%\eea 
%\end{widetext}

\section{Exponential waiting time distribution}\label{sec:exp}
%{Velocity as overdamped RTP}

In this section we consider the scenario where $\sigma(t)$ reverses sign with a constant rate $\beta$, resulting in an exponential waiting time distribution 
\begin{align}
\rho(\tau) = \beta e^{-\beta \tau}. \label{eq:rho_exp}
\end{align}
Consequently, the dichotomous acceleration develops an exponentially decaying auto-correlation,
\begin{align}
\la \sigma(t) \sigma(t') \ra = e^{-2\beta |t-t'|}.\label{eq:exp:corr}
\end{align}
At times $t \gg \beta^{-1},$ the auto-correlation decays, and we expect that the DAP in this regime will show a behaviour similar to the RAP, with a Gaussian position distribution and a variance which grows as $t^3.$ On the other hand, at short-times, we expect strong signatures of the activity. These features are already visible in the moments of the position distribution which we compute first.

\subsection{Moments of the position}

We compute the first two position moments using the Laplace transforms computed earlier in Eqs. \eref{eq:xav_laplace}-\eref{eq:x2av_laplace}. To this end, we need the Laplace transform of the exponential waiting time distribution \eref{eq:rho_exp}, which is given by,
\bea
g(s) = \frac{\beta}{\beta + s}.
\eea
From Eq.~\eqref{eq:xav_laplace} we then have,
\bea 
m_1(s) = \frac{a_0 \la \sigma_1 \ra}{s^2(s+2\beta)},
\eea 
which, upon inverting the Laplace transform, yields the average position of the particle,
\bea
\la x(t) \ra = \frac{a_0 \la \sigma_1 \ra }{4 \beta^2}[2 \beta t -1 + e^{-2\beta t}].
\eea
 Obviously, the average position vanishes if the  initial acceleration $\sigma_1 = \pm 1$ with equal probability. 

\begin{figure}
\centering \includegraphics[width=0.65\hsize]{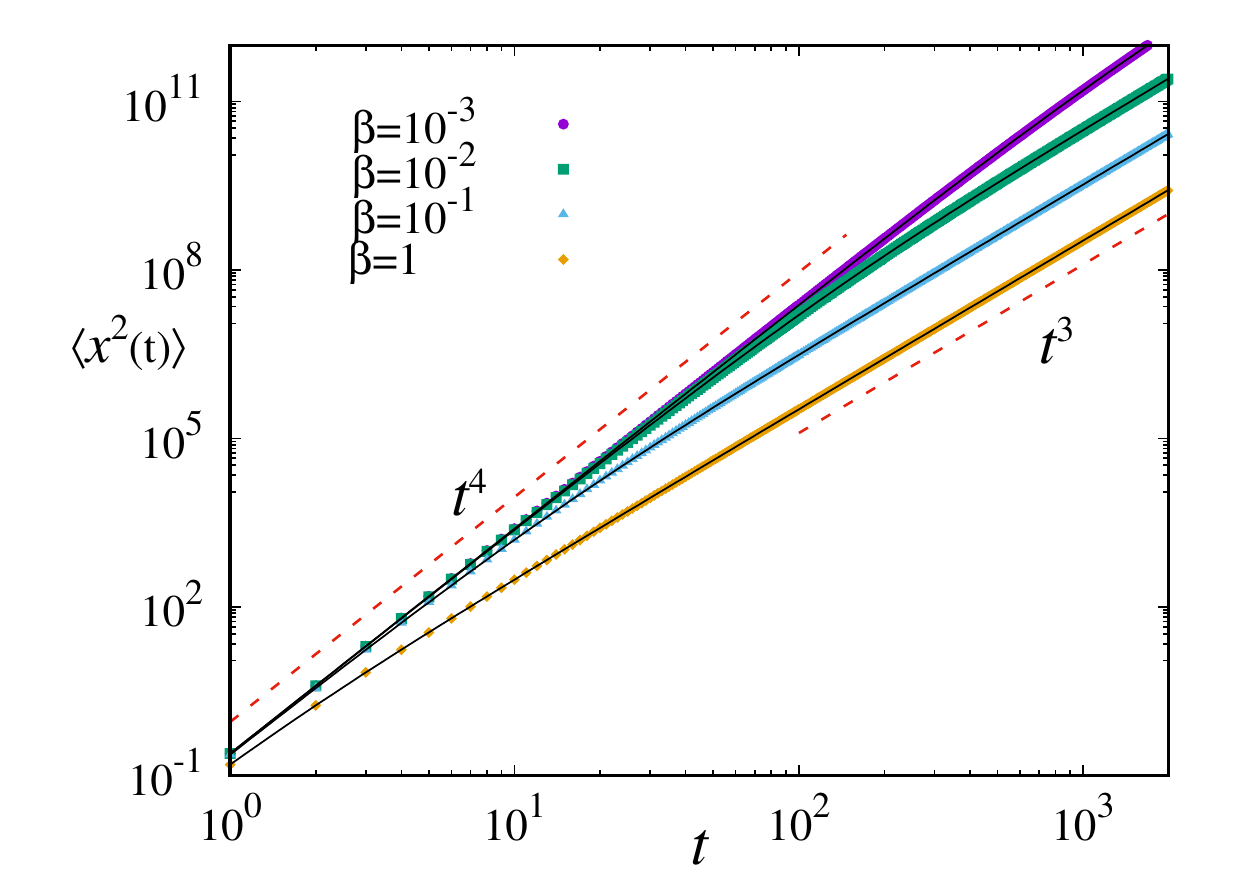}
\caption{Exponential waiting time: Second moment of the displacement $\la x^2(t) \ra$ as a function of time $t$ for different values of the flip rate $\beta.$ The symbols denote the data obtained from numerical simulations, while the solid black lines denote the theoretical prediction Eq.~\eqref{eq:expx2}. The dashed red lines indicate the asymptotic behaviors in the short-time and long-time regimes.}
\end{figure}

The Laplace transform of the second moment can be obtained from Eq.~\eqref{eq:x2av_laplace},
\bea
m_2(s)= 2 a_0^2 \frac{(4\beta+ 3 s)}{s^4(2\beta+s)^2},
\eea
which,  in turn,  gives,
\bea
\la x^2(t) \ra = \frac {a_0^2}{3 \beta} t^3 - \frac {a_0^2}{4 \beta^2} t^2 +\frac {a_0^2} {8\beta^4}[1- (1+2 \beta t)e^{-2 \beta t}].\label{eq:expx2}
\eea
Note that the second moment does not depend on the initial sign of the acceleration.
The short-time and long-time behaviours can be extracted easily from the above equation,
\bea
\la x^2(t) \ra = \left \{ 
\begin{array}{cc}
\frac{a_0^2}4 t^4 + {\cal O}(t^5) & \text{for} ~~ t \ll \beta^{-1}\cr
\frac {a_0^2}{3 \beta} t^3 + {\cal O}(t^2) & \text{for} ~~ t \gg \beta^{-1}.
\end{array}
\right.
\eea
This can be understood by noting that, the $v$-process is mathematically identical to an overdamped RTP, and shows a ballistic behaviour at short times with $\la v^2(t)\ra \sim t^2$ leading to the $\sim t^4$ growth at short times. At late times $t \gg \beta^{-1}$, the dichotomous noise emulates a delta-correlated white noise and the behaviour of the mean-squared displacement is consistent  with that of a RAP $\sim t^3$ which has already been shown in Ref.~\cite{Dean2021}. Note that, this $t^3$ growth, sometimes referred to as the Richardson-Obukhov law, is also observed in the context of anomalous diffusion of cold atoms~\cite{Barkai2014}.

\subsection{Position distributions}

The position distribution of the DAP is given by $P(x,t) =\sum_{\sigma=\pm 1}\int dv {\cal P}_\sigma(x,v,t)$ where ${\cal P}_\sigma(x,v,t)$ denotes the probability that the particle is at position $x$ with velocity $v$ and acceleration $\sigma$ at time $t$. For the exponential waiting time case, it is straightforward to write the Fokker-Planck equation  corresponding to the  Langevin equations \eref{eq:xv},
\bea
 \frac{\partial {\cal P}_{\sigma}}{\partial t} = - v \frac{\partial {\cal P}_{\sigma}} {\partial x} - \sigma \frac{\partial {\cal P}_{\sigma}} {\partial v} - \beta ({\cal P}_{\sigma} - {\cal P}_{-\sigma}).\label{FP_exp}
\eea
However, this set of coupled differential equations is difficult to solve exactly. Instead, we use the trajectory based approach developed in Sec.~\ref{sec:formalism} to investigate the asymptotic behaviour of the position distribution in the short-time ($t \ll \beta^{-1}$) and long-time regimes.

\begin{figure}
\includegraphics[width=\hsize]{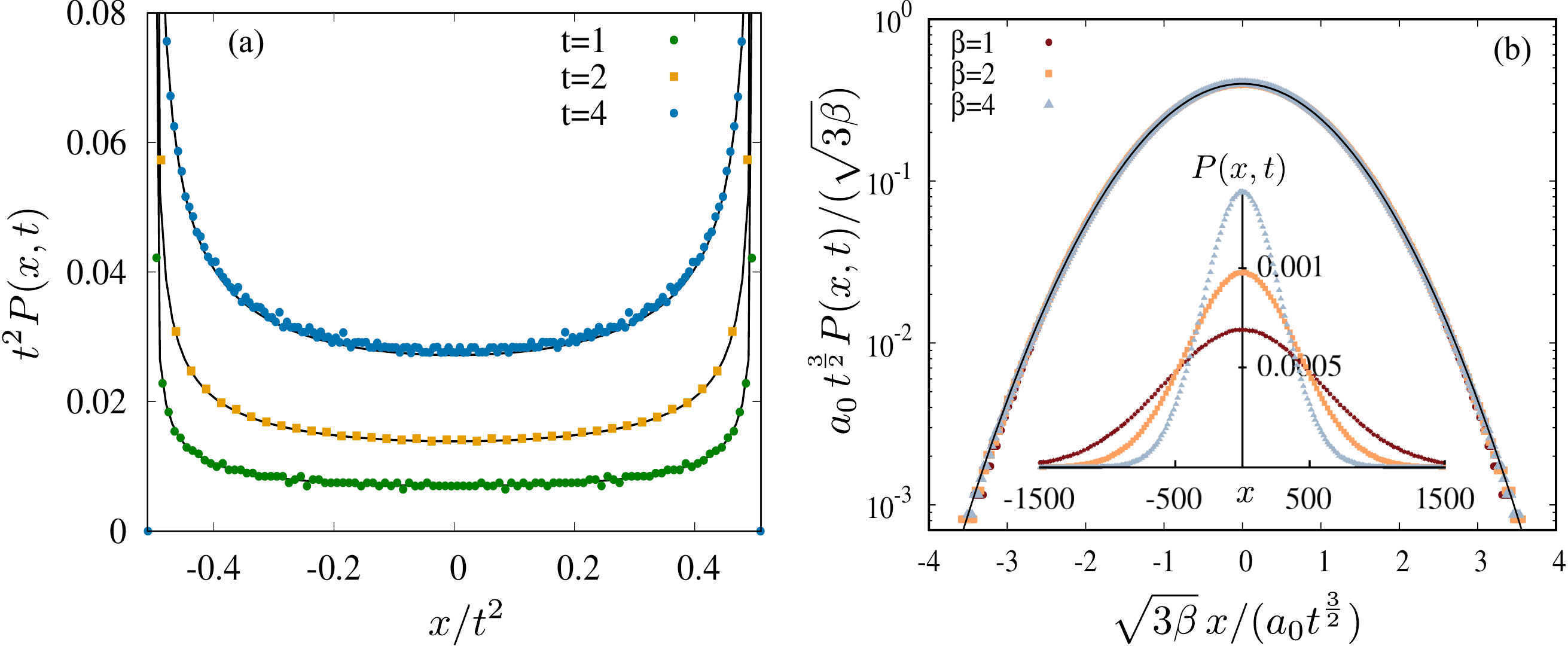}
\caption {Position distribution $P(x,t)$ of the DAP with exponential waiting-time distribution: (a) Plot of  $P(x,t)$ for different values of $t$ in the short-time regime $t \ll \beta^{-1}$, for a fixed $\beta=0.01$. The solid black lines indicate the analytical prediction from the perturbative procedure [see \eref{eq:Pxt_gen}] truncated at $n=2$, while the symbols indicate the data from numerical simulations. We have not shown the $\delta$-peaks at $x= \pm x_0$ for better visibility.  (b) Plot of $P(x,t)$ in the long-time regime for different values of $\beta$ at $t=100$. The main figure shows the scaling collapse of the data obtained from numerical simulations (symbols) with the predicted Gaussian scaling function (solid  line) [see \eref{exp:longtime}]. The inset shows the corresponding unscaled data. We have taken $a_0=1$ for both the panels.}
\label{f:exppdf}
\end{figure}

\subsubsection{Short-time regime:}

In the short-time regime, i.e., for $t \ll \beta^{-1}$, the typical number of acceleration reversal events are small. Consequently, we can perturbatively obtain the position distribution in this regime, by computing $P_{n,\sigma_1}(x,t)$ in \eref{eq:Pxt_gen} starting from $n=0$ and truncating at a finite (small) $n$. 

For $n=0,$ i.e., when there is no reversal of acceleration during time $t$, the particle moves deterministically with a constant acceleration $a_0\sigma_1$ and we have,
\begin{align}
P_{0,+}(x,t)=e^{-\beta t}~\delta(x-x_0),\quad \text{and}\quad  P_{0,-}(x,t)=e^{-\beta t}~\delta(x+x_0),\quad\text{with }x_0=\frac{a_0t^2}2.
\end{align}
Thus, following \eref{eq:Pxt_gen}, the contribution to the position distribution from the $n=0$ trajectories is given by,
\bea
P_0(x,t)=\frac 12e^{-\beta t}\left[\delta(x-x_0)+\delta(x+x_0)\right].
\eea
Clearly, because of the deterministic nature of the $n=0$ trajectories, the particle does not populate the region $x\in (-x_0,x_0)$. This situation changes when there are finite number of reversal events, starting with $n=1$. In this case, from \eref{eq:pnsigma_gen},
\bea
P_{1,+}(x,t)&=&\beta\,e^{-\beta t}\int_0^t d\tau_1\, \delta\left(x-a_0 \left[2t\tau_1-\tau_1^2-\frac{t^2}2\right]\right).
\eea
This integral can be easily computed using the properties of the delta-function integrand, which is non-zero only at $\tau_1^*=t-\sqrt{(x_0-x)/a_0}$ in the region $\tau_1\in[0,t]$. Then we have,
\bea
P_{1,+}(x,t)&=&\frac{\beta e^{-\beta t}}{2\sqrt{a_0(x_0 -x)}}.   
\eea
Using \eref{pnsymmetry} and \eref{eq:Pxt_gen}, we can immediately write the contribution of the $n=1$ trajectories to the position distribution,
 \bea
P_1(x,t) = \frac{\beta e^{-\beta t}}{4}   \bigg [\frac 1 {{\sqrt{a_0(x_0 -x)}}} +\frac 1 {{\sqrt{a_0(x_0 +x)}}} \bigg].
\eea
Evidently, $P_1(x,t)$ diverges algebraically near the boundaries $x=\pm x_0$.

 At the next order, i.e., for $n=2$, we get from \eref{eq:pnsigma_gen},
\begin{align}
 P_{2,+}(x,t)= \frac{\beta^2}2 e^{-\beta t}\int_0^t d\tau_1 \int_0^{t-\tau_1}d\tau_2 ~\delta \Bigg(x-a_0 \Big[t_2^2 -2 \tau_2(t-\tau_1)+\frac {t^2}2\Big] \Bigg).
\end{align}
Within the domain of the inner integral $t_2\in[0,t-\tau_1]$, the integrand is non-zero only at $t_2^*=t-\tau_1-\sqrt{\frac{x+x_0}{a_0}-\tau_1(2t-\tau_1)}$. Thus, we get,
\begin{align}
P_{2,+}(x,t)&= \frac{\beta^2}{4\sqrt{a_0}} e^{-\beta t}\int_0^t  \frac{d\tau_1}{\sqrt{x+x_0-a_0\tau_1(2t-\tau_1)}}=\frac{\beta^2}{8a_0} e^{-\beta t}\log\Bigg[\frac{\sqrt{x+x_0}+\sqrt{a_0}t}{\sqrt{x+x_0}-\sqrt{a_0}t}\Bigg].
\end{align}
$P_{2,-}(x,t)$ can be obtained from the above equation  using \eref{pnsymmetry}. Adding the two and simplifying, we get the total contribution from the trajectories with two reversal events,
\begin{align}
P_2(x,t) = \frac {\beta^2\,e^{-\beta t}}{4 a_0} \left[ \log\left(\frac{\sqrt{a_0}\, t +  \sqrt{x_0+x}}{\sqrt{x_0-x}} \right) + \log\left(\frac{\sqrt{a_0}\, t +  \sqrt{x_0-x}}{\sqrt{x_0+x}} \right) \right].\label{eq:p2exp}
\end{align}
The contributions from higher number of reversal events can also be computed systematically, following the same procedure.
However, keeping upto $n=2$ already provides a reasonably good approximation of the short-time distribution. This is illustrated in~\fref{f:exppdf}(a), which compares the analytical prediction upto $n=2$ with the short-time $P(x,t)$ measured from numerical simulations.

\subsubsection{Long-time regime:}

At times $t\gg\beta^{-1}$, the typical number of reversal events are large and the dichotomous noise $\sigma(t)$ emulates a Gaussian white noise~\cite{Santra2021} with strength $1/(2\beta)$ [see \eqref{eq:exp:corr}]. Thus, in this regime, the typical position fluctuations of DAP follow,
\begin{align}
\ddot{x}(t)=\frac{a_0}{\sqrt \beta}\,\eta(t),\quad \text{with }\la\eta(t)\ra=0~~ \text{and }\la\eta(t)\eta(t')\ra=\delta(t-t').
\end{align}
This is nothing but a random acceleration process [see \eref{eq:rap1}]. The corresponding position distribution is a Gaussian with a super-diffusive scaling form,
\begin{align}
P(x,t)=\frac{\sqrt{3\beta}}{a_0t^{3/2}}G\left( \frac{\sqrt{3\beta}}{a_0t^{3/2}}x\right),\quad\text{with}~~
G(z)=\frac{1}{\sqrt{2\pi}}e^{-z^2/2}.\label{exp:longtime}
\end{align}
This is illustrated in \fref{f:exppdf}(b), where the scaling collapse of the long time distribution is shown for different values of $\beta$. 

Note that, though \eref{exp:longtime} describes the typical fluctuations very well, deviations from the Gaussian form are expected at the tails, i.e., for $x\sim O(t^2)$. The nature of these atypical fluctuations has been studied in details in~\cite{Dean2021}, where the large deviation function and the long time behavior of the higher moments have been derived in the context of a generalized run-and-tumble process.

\section{Power-law waiting time distribution}\label{sec:plaw}

In this section we consider the case where the waiting time is drawn from a power-law distribution,
\bea 
\rho(\tau) = \left \{ 
\begin{split}
 &0  ~~~~~~~\text{for}~~ \tau < \tau_0,  \cr 
 &\frac{\alpha\, \tau_0^\alpha}{\tau^{1+ \alpha}}~~\text{for}~~ \tau \ge  \tau_0.
\end{split}
\right.
\label{plawdist}
\eea 
where $\tau_0$ denotes the minimum waiting time, and the exponent $\alpha$ characterises the shape of the distribution. Smaller values of $\alpha$ indicates a slower decay of $\rho(\tau)$, and correspondingly, the typical waiting time is larger for smaller $\alpha$. In fact, the average waiting time $\la\tau\ra$ diverges for $\alpha\leq 1$ and $\la\tau\ra=\alpha\tau_0/(\alpha-1)$ for $\alpha>1.$ Additionally, the second moment   $\la\tau^2\ra$ diverges for $\alpha\leq 2$ and $\la\tau^2\ra=\alpha\tau_0^2/(\alpha-2)$ for $\alpha>2.$ This is in stark contrast with the exponential waiting time distribution, where all the moments are finite for all parameter values. 

The residual time distribution corresponding to \eref{plawdist} also decays algebraically,
\bea 
f(\tau) = \left \{ 
\begin{split}
& 1  ~~~~~~~~~~\text{for}~~ \tau < \tau_0,  \cr 
 &\left(\frac {\tau_0} \tau \right)^{\alpha }  ~~\text{for}~~ \tau \ge  \tau_0.
\end{split}
\right.
\eea 
The nature of the position fluctuations for this power-law waiting time distribution is expected to be  drastically different from that corresponding to the exponential waiting time distribution. In the following, we investigate the case of power-law waiting times by studying the position moments and distribution. 

\subsection{Moments}
The position moments are the primary indicators of the non-Markovian nature of the waiting time distribution $\rho(\tau)$. The trajectory based approach introduced in Sec.~\ref{sec:formalism} can be directly used to compute the first two moments of $x(t)$. To this end, we first note that the Laplace transform of $\rho(\tau)$ in \eref{plawdist} is given by,
\bea 
g(s) = \int_{\tau_0} ^\infty dt~ e^{-st} \rho(t) = \alpha (s \tau_0)^\alpha \Gamma(-\alpha, s\tau_0),
\eea 
where $\Gamma(\nu,w)$ denotes the upper incomplete Gamma function~\cite{dlmf}.
The Laplace transform $m_1(s)$ of the average position $\la x(t)\ra$ can then be obtained using the above equation in  \eref{eq:xav_laplace}. While it is hard to invert the Laplace transform in \eref{eq:xav_laplace} exactly, one can use the small $s$ and large $s$ behaviour of $m_1(s)$ to get the asymptotic behaviour of $\la x(t) \ra$ in long-time and short-time limits, respectively. For this purpose, we require the asymptotic behaviors of $g(s)$. For $s\tau_0\ll 1$,
\bea
g(s)=\begin{cases} \alpha (s\tau_0)^{\alpha}\Gamma(-\alpha)+1-\frac{\alpha}{\alpha-1}\,s\tau_0+O\left[(s\tau_0)^2\right] \quad\text{for }\alpha\neq 1,\\
1+s\tau_0\left[\log(s\tau_0)+E_\gamma-1\right]+(s\tau_0)^2\qquad~~\text{for }\alpha=1.
\end{cases}
\label{plawgs_asymptotes1}
\eea
On the other hand, for $s\tau_0\gg 1$,
\bea
g(s)=e^{-s\tau_0}\left(\frac{\alpha}{s\tau_0}+O\left[(s\tau_0)^{-2}\right]\right).
\label{plawgs_asymptotes2}
\eea
The large $s$ ($s>\tau_0^{-1}$) behavior of $m_1(s)$ is now obtained by using \eref{plawgs_asymptotes2} in \eref{eq:xav_laplace},
\bea
m_1(s)\simeq \frac{a_0\la\sigma_1\ra}{s^3},
\eea
which leads to a super-ballisitic behavior in the short-time regime $t<\tau_0$,
\bea
\la x(t)\ra=\frac{a_0 \la\sigma_1\ra}{2}t^2. 
\eea
Physically, this regime corresponds to the scenario where $\sigma$ has not flipped at all (recall that, $\rho(\tau)=0$ for $t<\tau_0$) and the particle moves with a constant acceleration. 

Next, we investigate the long-time behavior of $\la x(t)\ra$. For $s\ll\tau_0^{-1}$, we have from \eref{plawgs_asymptotes1} and \eref{eq:xav_laplace},
\bea 
m_1(s) \simeq \frac{a_0 \la \sigma_1 \ra}{2}\left\{
\begin{split}
\frac{\tau_0^\alpha \Gamma(-\alpha)}{s^{3-\alpha}}\quad&\text{for }\alpha<1\\
\frac{\tau_0\left[1-E_\gamma-\log(s\tau_0)\right]}{2s^2} \quad&\text{for }\alpha=1\\
\frac {\alpha \tau_0}{(\alpha-1)s^2}  \quad&\text{for }\alpha>1,
\end{split}\right.
\eea
where $E_{\gamma}\approx 0.5772$ denotes the Euler-Mascheroni constant. This leads to a non-trivial $\alpha$-dependent temporal growth of the average position for $t \gg\tau_0,$
\bea
\la x(t)\ra \simeq \frac{a_0 \la \sigma_1 \ra}{2}\left\{ 
\begin{split}
   \frac{\tau_0^\alpha \Gamma(-\alpha)}{\Gamma(3-\alpha)}~ t^{2-\alpha} &\qquad\qquad\alpha<1\cr
   \tau_0 t \log\Big(\frac t {t_0}\Big) & \qquad\qquad \alpha =1 \cr 
   \frac{\alpha \tau_0}{(\alpha -1)}~ t \quad & \qquad\qquad \alpha>1.
\end{split}
\right.
\label{plaw_m1(t)}
\eea

\begin{figure}
\includegraphics[width=\hsize]{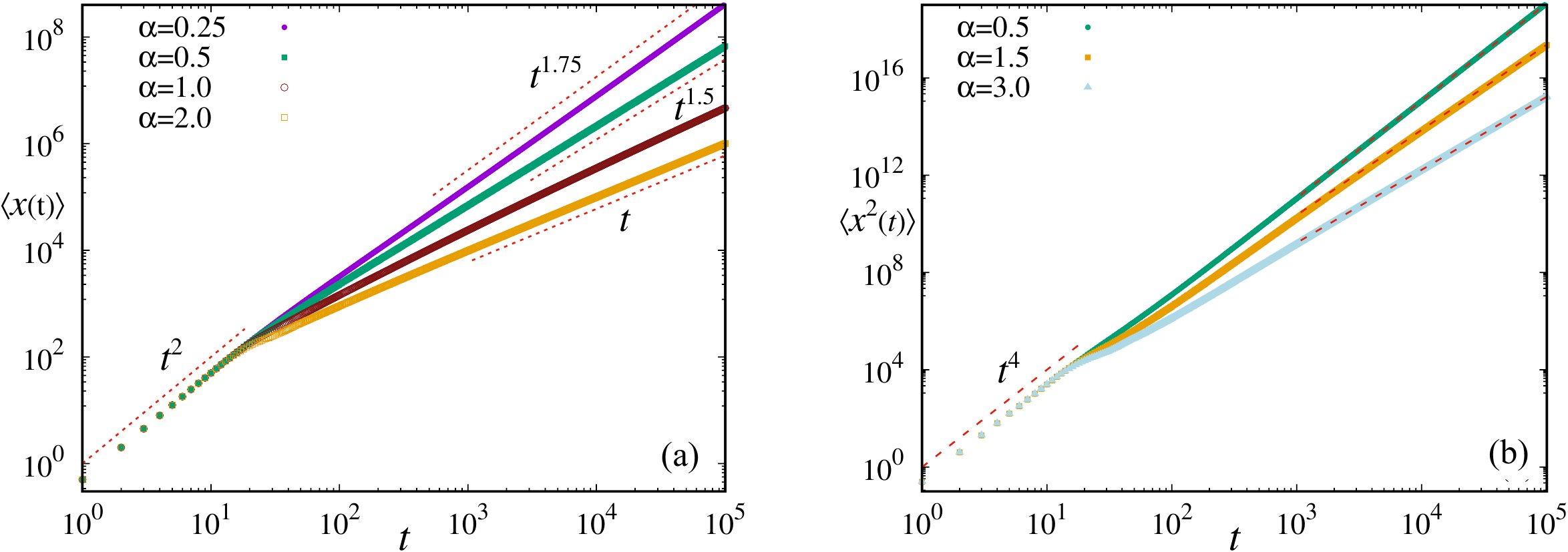}
\caption{Position moments for power-law waiting time distribution: Plot of (a) $\la x(t) \ra$ and (b) $\la x^2(t) \ra$ vs $t$ for different values of $\alpha$. The symbols indicate the data from numerical simulations and the dashed lines indicate the analytically predicted algebraic growth. Here we have used $a_0=1$, $\tau_0=10$ and initial acceleration sign $\sigma_1=1$.}
\label{moments_plaw}
\end{figure}

Clearly, when $\sigma_1=\pm 1$ with equal probability, i.e., the initial acceleration direction is equally likely to be along $+x$ and $-x$, the average displacement vanishes. The predictions in \eref{plaw_m1(t)} are compared with numerical simulations in \fref{moments_plaw}(a) for $\la\sigma_1\ra=1$ which shows an excellent agreement.

To compute the second moment, we proceed similarly using \eref{eq:x2av_laplace}. The behavior of $\la x^2(t)\ra$ at short-times can be obtained from the leading order behavior of $m_2(s)$ for $s\gg\tau_0^{-1}$,
\bea
m_2(s)\simeq 6a_0^2\,s^{-5},
\eea
which leads to,
\bea
\la x(t)^2\ra=\frac{a_0^2 }{4}t^4, \quad\text{  for  }t<\tau_0.
\eea
Physically, this again corresponds to a deterministic motion with a constant acceleration $a_0$ since $\sigma$ does not flip for $t<\tau_0$. 

 For $s\ll\tau_0^{-1}$, the leading order behavior of the Laplace transform of the second position moment $m_2(s)$ is given by,
\bea
m_2(s)\simeq a_0^2\left\{ 
\begin{split}
    &(6-\alpha)(1-\alpha)~s^{-5}\quad \qquad \qquad \qquad ~\alpha < 1\cr
    &\frac{\tau_0^{\alpha -1}}{\alpha }\Gamma(2-\alpha)(6-\alpha)(\alpha-1) \,s^{\alpha-6}\quad   1< \alpha <2\cr 
 &\frac{2\tau_0}{(\alpha-2)(\alpha -1)}\, s^{-4}\quad \qquad \qquad \qquad~ \alpha > 2 .  
\end{split}
\right.
\eea
These can be easily inverted to obtain the long-time behavior of the second moment of the position,
\bea
\la x^2(t) \ra \simeq a_0^2\left\{ 
\begin{split}
    &\frac{(6-\alpha)(1-\alpha)}{24}\, t^4,\quad \qquad \qquad \qquad ~\alpha < 1\cr
    &\frac{\Gamma(2-\alpha)(6-\alpha)(\alpha-1) \tau_0^{\alpha -1}}{\alpha \Gamma(6-\alpha)}\, t^{5-\alpha}\quad  1< \alpha <2\cr 
 &\frac{\tau_0}{3(\alpha-2)(\alpha -1)}\, t^3,\quad \qquad \qquad \qquad~ \alpha > 2 .  
\end{split}
\right.\label{plaw:2ndmom}
\eea
The above predictions are compared with numerical simulations in \fref{moments_plaw}(b) which show excellent agreement.

Note that, the above results are not expected to be valid at the special points $\alpha=1,\,2$ 
since the coefficients of $t^4$ (for $\alpha=1$) vanishes at $\alpha=1$, and the coefficients of $t^3$ diverge as $\alpha=2$ is approached from above and below. Consequently, non-trivial corrections are expected at $\alpha=1,\,2$, which are very hard to compute.

It is noteworthy that, a wide range of anomalous temporal growth of position variance is also observed in the context of L\'evy walks, where the particle displacement in each step is coupled to the duration of the step~\cite{Barkai1998,Barkai2018}. However, the growth exponents in those cases depend on the specific forms of the couplings.

\subsection{Position distribution}
The non-Markovian evolution of the acceleration implies that one cannot write a standard Fokker-Planck equation like \eref{FP_exp} for the position distribution. However, as before, we use the trajectory based approach to investigate the short-time and long-time distribution of the position separately. For simplicity, we consider the particle starting from origin with initial acceleration positive ($\sigma_1=1$) and negative ($\sigma_1=-1$) with equal probability.

\subsubsection{Short-time regime:}
We start from the formal expression \eref{eq:Pxt_gen} for the position distribution $P(x,t)$ which considers contributions from trajectories with $n$ number of acceleration reversals, for all possible values of $n$. It should be noted that, since $\rho(\tau)=0$ for $\tau<\tau_0$, for any finite $t\leq m\tau_0;\, m=1,2,\dotsc$, the exact position distribution can be computed by considering contributions from trajectories with $n\leq m$ flips. Thus, to understand the short-time behavior of the position distribution, it suffices to calculate the contributions from trajectories with small number of acceleration reversing events. In the following, we compute explicitly $P_{n,\sigma_1}(x,t)$ for $n=0,\,1$ and show that the shape of the distribution differs qualitatively from the exponential waiting time scenario, the position density varying non-monotonically as one moves farther away from the origin.

%For $n=0$, 
%there are no acceleration reversals during the time interval $[0,t]$ the particle travels with a constant acceleration $a_0\sigma_1$. Correspondingly, we 
The first contribution comes from the $n=0$ term which corresponds to a deterministic motion of the particle with a constant acceleration $a_0\sigma_1$. In this case, we have from \eqref{eq:Pxt_gen}, 
\bea
 \fl\qquad P_{0,\sigma_1}(x,t)&=&f(t)\,\delta\Bigg(x-\frac{a_0\sigma_1t^2}{2}\Bigg)= \left \{ \begin{split}
&\delta\left(x_0-x\right) \qquad ~~\text{for}~~ t < \tau_0  \cr 
& \left(\frac {\tau_0} t \right)^{\alpha } \delta\left(x_0 -x\right) ~~\text{for}~~ t \ge  \tau_0,
\end{split}
\right.
\label{plaw:p01}
\eea
where $x_0=a_0t^2/2$.
Similar to the exponential waiting time scenario, this gives rise to two delta-peaks at $x=\pm x_0$ for $\sigma_1=\pm 1$. However, the weight $f(t)$  of such trajectories are different for $t<\tau_0$ and $t\geq\tau_0$. In fact, for  $t<\tau_0$, $f(t)=1$ and \eref{plaw:p01} gives the exact position distribution,
\begin{align}
P(x,t)=\frac12\Big(P_{0,+}(x,t)+P_{0,-}(x,t)\Big).
\end{align}

\begin{figure}[t]
\centering\includegraphics[width=\hsize]{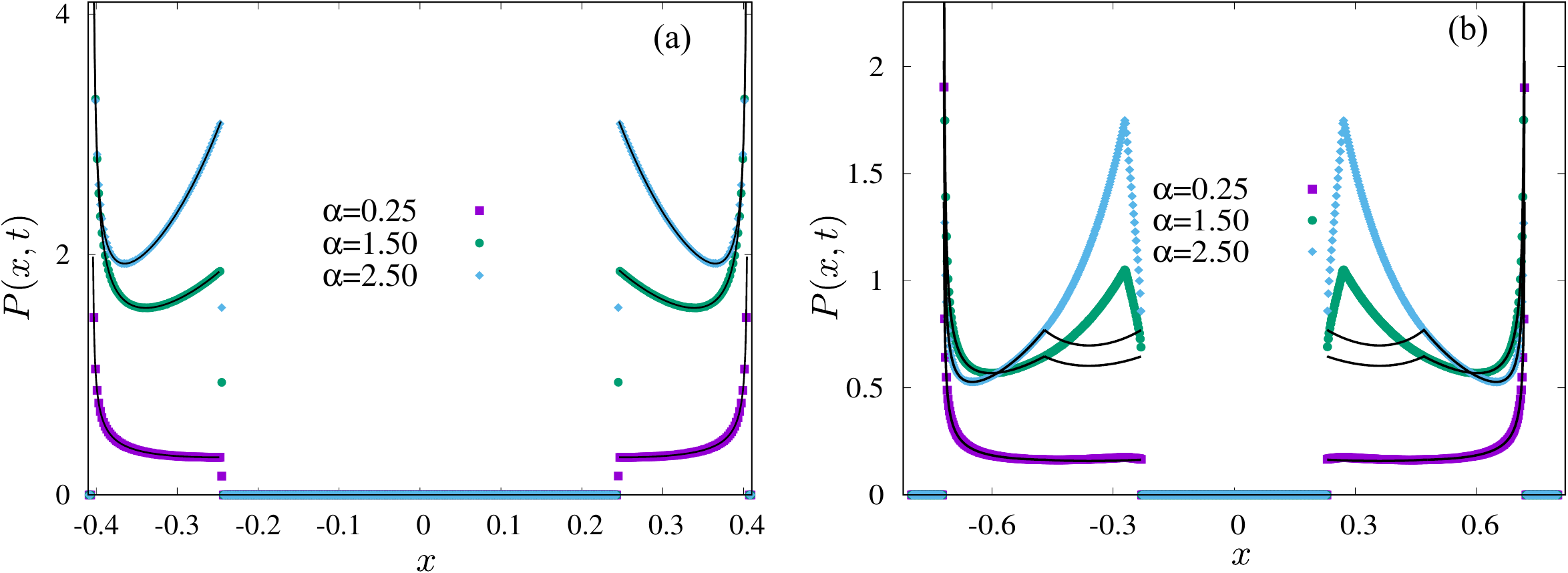}
\caption{Power-law waiting time: Short-time position distribution $P(x,t)$ for $\tau_0=0.5$ and different values of $\alpha$. The data obtained from numerical simulations (symbols) in the regimes $t<2\tau_0$ with $t=0.9$  (a)  and $t>2\tau_0$ with $t=1.2$ (b) are compared with the analytical predictions from the perturbative expansion evaluated up to $n=1$ [see \eref{plaw:trunc}]. Here $a_0=1$.}\label{f:plaw_st}
\end{figure}

For $t>\tau_0$, the contributions from $n>0$ become important. From \eqref{eq:Pxt_gen}, we have for $n=1$,
\begin{align}
P_{1,\sigma}(x,t)=\int_{\tau_0}^t \id \tau_1\, \rho(\tau_1) f(t-\tau_1)\, \delta\left(x-a_0\sigma_1 \left[2t\tau_1-\tau_1^2-\frac{t^2}2\right]\right).
\label{plaw:oneflip}
\end{align}
To compute the above integral explicitly, we need to distinguish between the two scenarios $t<2\tau_0$ and $t\geq 2\tau_0$. Let us consider the former case first. Since the first flip can occur earliest at $\tau_1=\tau_0$, the interval $(t-\tau_1)$ is always smaller than $\tau_0$, implying $f(t-\tau_1)=1$ in this regime. Using this, we finally get,
\begin{align}
P_{1,+}(x,t)=&\frac{\rho(t^*(x))}{\sqrt{4a_0(x_0-x)}}\Theta(x_0-x)\Theta(x-x_1),\\
P_{1,-}(x,t)=P_{1,+}(-x,t)=&\frac{\rho(t^*(-x))}{\sqrt{4a_0(x_0+x)}}\Theta(x_0+x)\Theta(-x-x_1),
\end{align}
where $t^*(x)=t-\sqrt{\frac{x_0-x}{a_0}}$ and $x_1=a_0(2\tau_0 t-t^2/2-\tau_0^2)$. Again, it is important to note that, since there cannot be more than one-flip in this regime $t<2\tau_0$, the exact position distribution is given by,
\begin{align}
P(x,t)=\frac12\sum_{n=0,1}\left(P_{n,+}(x,t)+P_{n,-}(x,t)\right).\label{plaw:trunc}
\end{align}
Figure~\ref{f:plaw_st}(a) shows a plot of the position distribution in the $t<2\tau_0$ regime, which validates the above prediction.

If, however, $t>2\tau_0$, the interval $t-\tau_1$ can be both smaller or larger than $\tau_0$ and accordingly $f(t-\tau_1)$ in \eqref{plaw:oneflip} equals $1$ or $\Big(\tau_0/(t-t_1)\Big)^\alpha$. Performing the integral in \eref{plaw:oneflip} using the appropriate form of $f(t)$, we get, 
\begin{align}
\fl \qquad \qquad \quad P_{1,+}(x,t) = \frac{\rho(t^*(x))}{\sqrt{4a_0(x_0-x)}}\Bigg[
\Theta[x_2-x]\Theta[x-x_1]+ 
\frac{(\sqrt a_0\tau_0)^{\alpha }}{(x_0-x)^{\alpha/2 }}
 \Theta[x-x_2]\Theta[x_0-x]
\Bigg].
\end{align}
Then, the total contribution to the position distribution $P_1(x,t)$ coming from $n=1$ is given by,
\begin{align}
P_{1}(x,t)=\frac12\Big(P_{1,+}(x,t)+P_{1,-}(x,t)\Big)=\frac12\Big(P_{1,+}(x,t)+P_{1,+}(-x,t)\Big).
\end{align}
Clearly, the distribution obtained by calculating the contributions from the zero and one flip terms are not exact for $t>2\tau_0$, since in this time regime, there can be more than one flips. Numerical simulations suggest that, the deviation of the exact distribution with this approximate distribution, truncated at one-flip term, increases with increase in $\alpha$. This is expected since the typical waiting time decreases with increase in $\alpha$, thus, increasing the contributions from trajectories with larger number of flips. This is illustrated in \fref{f:plaw_st}(b).

\subsubsection{Long-time regime:}
\begin{figure}
\centering \includegraphics[width=\hsize]{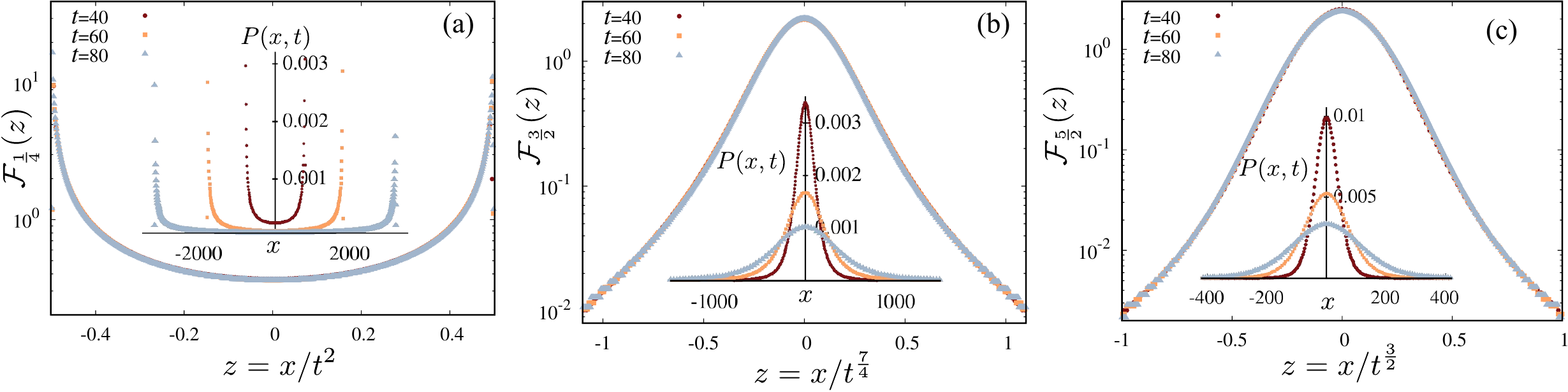}
\caption{Long time position distribution for power-law waiting times. The main figures show the scaling function ${\cal F}_{\alpha}(z)$ [see \eref{plaw_sc_lt}] as a function of $z=x/ t^{\phi(\alpha)}$ for different values of $t$ and (a) $\alpha=1/4$, (b) $\alpha=3/2$ and  (c) $\alpha=5/2$. The insets show the corresponding unscaled data. Here we have used $a_0=1$ and $\tau_0=0.1$. }\label{f:plaw_lt}
\end{figure}

In contrast to the exponential waiting time distribution, there is no simple argument to predict the position distribution in the long-time regime $t\gg \tau_0$ for all values of $\alpha$.  It is also difficult to analytically compute the position distribution in this regime using the trajectory-based approach. We use numerical simulations to investigate the long-time position distribution---\fref{f:plaw_lt} shows plots of the numerically measured $P(x,t)$ for $t\gg \tau_0$ for different values of $\alpha$, which clearly indicates that the shape of the shape of the distribution is qualitatively different for $\alpha<1$ and $\alpha>1$.

The asymptotic forms of the second moment of $P(x,t)$ given by \eref{plaw:2ndmom}  suggest how typical scales of the position fluctuations vary with time. Based on that, we propose a scaling form  for the typical position fluctuations, 
\begin{align}
P(x,t)=t^{-\phi(\alpha)}{\cal F}_\alpha\Bigg( \frac{x}{t^{\phi(\alpha)}}\Bigg)~~~~\text{with }\phi(\alpha)= 
\begin{cases}2~~&\text{for}~\alpha<1\\
\frac{5-\alpha}2~~&\text{for}~1<\alpha<2\\
\frac 32~~&\text{for}~\alpha>2,
\end{cases}
\label{plaw_sc_lt}
\end{align}
where ${\cal F}_{\alpha}(z)$ denotes the scaling function. Figure \ref{f:plaw_lt} (main plots) shows plots of the distribution of the scaled variable $z=x/t^{\phi(\alpha)}$ for three different values of $\alpha$, in the ranges $\alpha<1$, $1<\alpha<2$ and $\alpha>2$. Perfect collapses are observed in all the three cases, validating the scaling conjecture \eref{plaw_sc_lt}. Interestingly, the shape of the scaling function for $\alpha<1$ is very different than that for $\alpha>1$. For $\alpha<1$, the scaling function has peaks near $z=\pm a_0/2$ (i.e., $x=\pm a_0t^2/2)$. 
This can be qualitatively understood by remembering that for $\alpha<1$ the average waiting time between two consecutive reversals diverges. As a result of which the contribution to probability density from trajectories with zero or very few reversals is substantial leading to the peaks near $x=\pm a_0t^2/2$.
 On the other hand, for $\alpha>1$, the scaling function has a single peak around $z=0$, although the actual functional form appears to be different for $\alpha<2$ and $\alpha>2$.
 
\begin{figure}
\centering \includegraphics[width=\hsize]{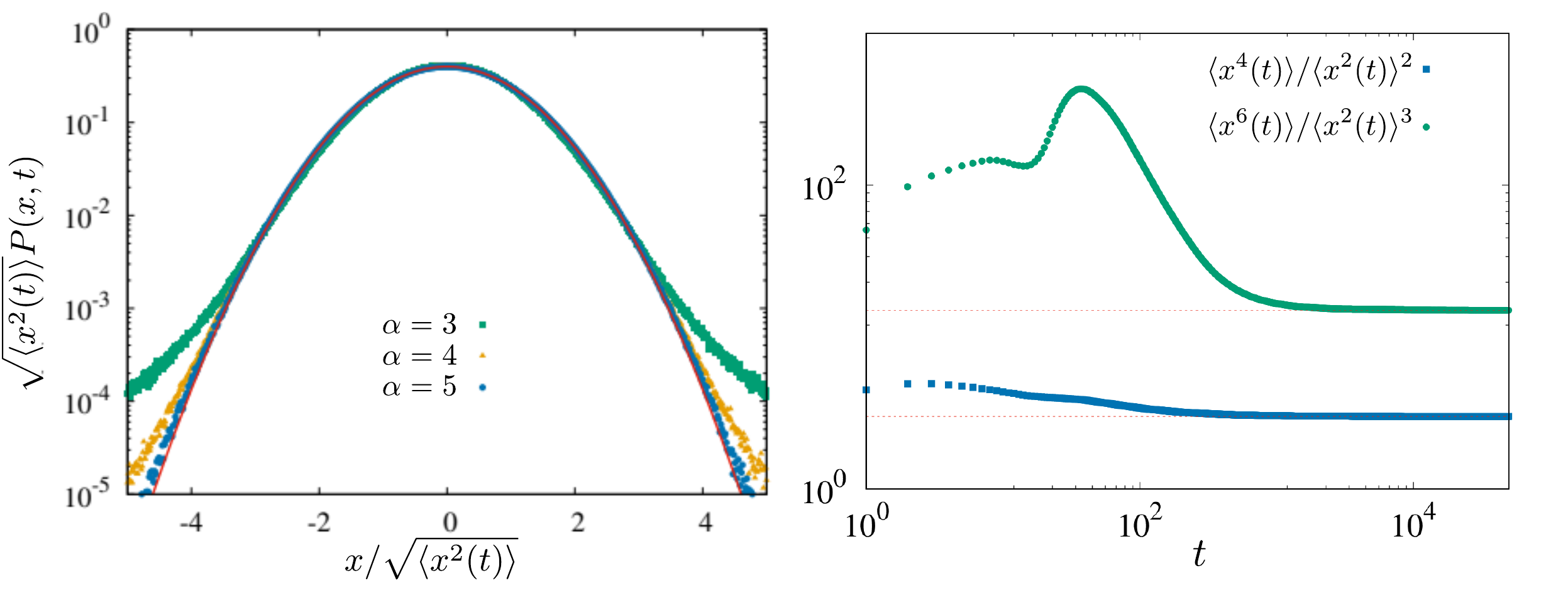}
\caption{Approach to Gaussian distribution at long-times for $\alpha>2$: (a) Plot of the scaled distribution for different $\alpha$. The symbols denote the position distribution as a function of the scaled variable $z=x/\sqrt{\la x^2(t)\ra}$ for different values of $\alpha$ and $t=500$. The solid red line denotes the standard normal distribution $e^{-z^2/2}/\sqrt{2\pi}$, which matches with the simulation data progressively better as $\alpha$ is increased. (b) Time evolution of the fourth and sixth position moments, obtained from numerical simulations, scaled with appropriate powers of the variance for $\alpha=4$ and $\tau_0=0.05$. The values of these quantities saturate to the predicted ones for a Gaussian distribution~\eref{eq:gauss-m}, indicated by the red dashed lines. } \label{plaw:lt-gaussian}
\end{figure} 

For $\alpha>2$, the waiting time distribution has finite mean and variance. In this case, the number of reversals are large in the long time regime and one can, using central limit theorem, heuristically argue that the typical position fluctuations should be Gaussian. This argument becomes more accurate as $\alpha$ is increased. This is shown in \fref{plaw:lt-gaussian} where the distribution of the scaled variable $x/\sqrt{\la x^2(t)\ra}$ for different values of $\alpha>2$ are compared with the standard normal distribution, which shows an increasingly better agreement as $\alpha$ is increased.
 To illustrate this further, we also measure higher moments of the position, which, for a Gaussian distribution, are simple functions of the variance. In particular, for a Gaussian distribution $g(z)=\exp(-\frac{z^2}{2\la z^2\ra})/\sqrt{2\pi}$, 
\begin{align}
\la z^4\ra=3\la z^2\ra^2,\qquad\la z^6\ra=15\la z^2\ra^3.
\label{eq:gauss-m}
\end{align}
Figure \ref{plaw:lt-gaussian}(b) shows the time evolutions of $\la x^4\ra/\la x^2\ra^2$ and $\la x^6\ra/\la x^2\ra^3$ of DAP for $\alpha=4$, which shows that the higher order position fluctuations become consistent with a Gaussian distribution as time increases.

\begin{figure}
 \includegraphics[width=\hsize]{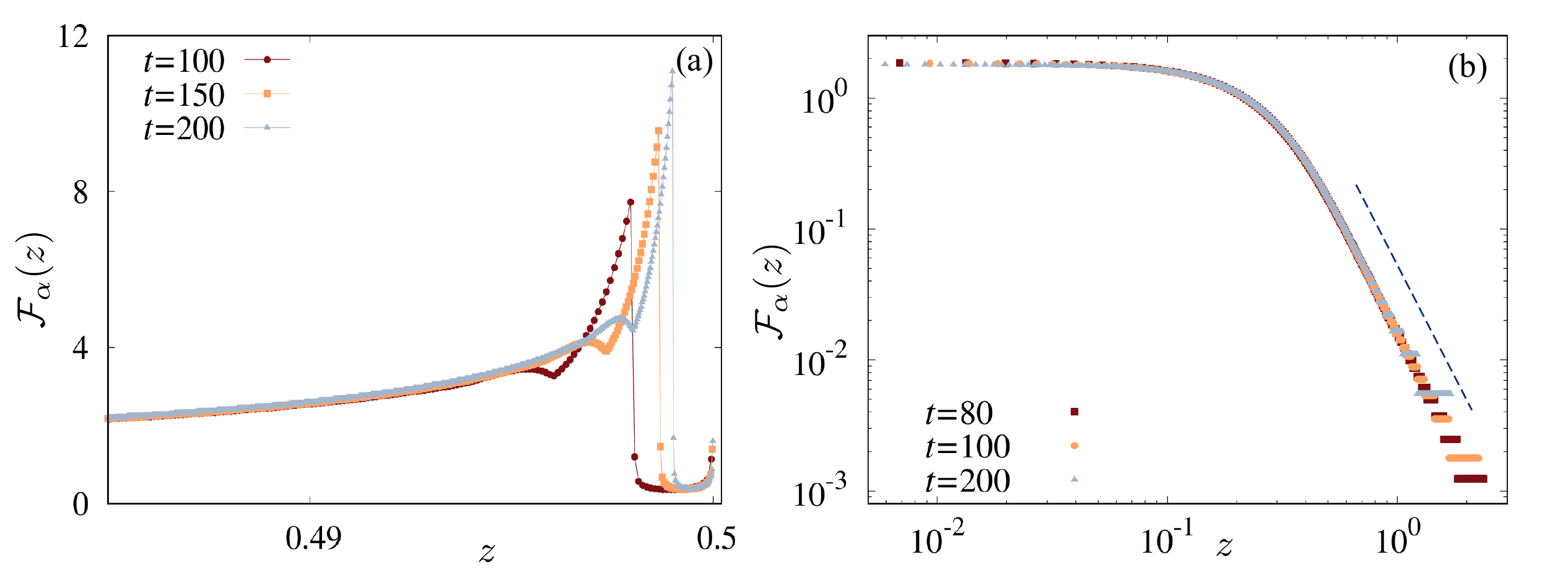}
 \caption{ Tails of the scaling function $\mathcal{F}_\alpha(z)$ at long times for $\alpha<1$ and $\alpha>1$. (a) The behavior of the scaling function for $\alpha=1/2$ for three different values of $t$, zoomed in near $z=1/2$. The absence of the collapse of the different curves near their peaks indicates that the predicted scaling form does not hold near the tails. (b) The behavior of the tails of $\mathcal{F}_\alpha(z)$ for $\alpha=1.75$. A perfect collapse indicates that the scaling form holds even at the tails, where it shows a power-law decay (indicated by dashed grey line).
 }
 \label{f:tails}
 \end{figure}

It is worthwhile to investigate the behavior of the tails of the scaling functions for $\alpha<2$. Figure~\ref{f:tails}(a) shows a plot of $\cal{F}_{\alpha}(z)$ for $\alpha=1/2$, zoomed in near the boundary $z=1/2$. Interestingly, a clear deviation from the scaling form is visible around the peak, which appears at some $z^*<1/2$. 
For $\alpha>1$, the scaling form \eref{plaw_sc_lt} holds true even at the tails, as shown in \fref{f:tails}(b). Moreover, it appears that the tails of the scaling functions decay algebraically, with a possibly $\alpha$-dependent exponent. However, we do not have any analytical prediction for this decay exponent.

\section{Conclusion}\label{sec:conc}

 We study the motion of a one-dimensional particle with dichotomous stochastic acceleration---the magnitude of the acceleration remains same, while its direction reverses stochastically after a random waiting-time. Such a dynamics leads to a generic non-Markovian evolution for the position and we develop a trajectory based approach to study it. We compute exactly the Laplace transform of the mean and the variance of the position for any arbitrary waiting-time distribution. In  particular, we focus on two contrasting scenarios: first, the acceleration reversal occurring  at a constant rate, which corresponds to an exponentially decaying waiting-time distribution, and second, a heavy-tailed waiting-time distribution, decaying algebraically. We show that, while the short-time behavior of the moments are independent of the specific waiting time distribution, the long time behavior is very different. In fact, for the power-law waiting time distribution with decay exponent $\alpha$, the particle shows a wide range of super-diffusive behavior where the variance of the position fluctuations grows with an $\alpha$ dependent exponent.
 
We also explore the behavior of the position distribution in the short-time and long-time regimes for the exponential and power-law waiting-time distributions. In both cases, the position distribution is bounded in the region $[-a_0 t^2/2,a_0 t^2/2]$, where $a_0$ is the magnitude of the acceleration. We find that, at short-times the position distribution algebraically diverges near the boundary for both cases, although the behavior near the origin shows much richer features for the power-law case. On the other hand, the long-time behavior of the position distribution turns out to be very different in the two cases. When the waiting time is distributed exponentially, the long-time position distribution is always Gaussian with the width scaling as $t^{3/2}$ (similar to RAP). For power-law waiting-time distributions, however, the position distribution has qualitatively different shapes depending on the exponent $\alpha$. Although the analytical forms of these distributions are not known, we show that the distributions can be expressed in terms of the scaled variable $x/t^{\phi(\alpha)}$, where $\phi(\alpha)=2$, $(5-\alpha)/2,$ and $3/2$ for $\alpha<1,~1<\alpha\leq 2$ and $\alpha>2$ respectively.

For exponential waiting time distribution, the DAP shows an interesting large deviation behavior~\cite{Dean2021}. An obvious question is how the large deviation function changes if the waiting time distribution is power-law---in particular, if one can apply the big jump principle previously used in the context of L\'evy walks~\cite{Burioni2019}.
The first passage properties of random acceleration processes have been of interest for a long-time~\cite{Burkhardt2014}. It would also be intriguing to study the first-passage behavior of the dichotomous acceleration process, particularly for the power-law waiting time distribution. Another interesting question is how the presence of dissipation and thermal noise affects the behavior of DAP, which would relate to the behavior of inertial active particles~\cite{Lowen2020inertial}.

\appendix
\addtocontents{toc}{\fixappendix}

\section{Computation of moments}\label{app:moments}

In this section we provide details about the explicit computation of the Laplace transform of the first two moments of DAP. 

\subsection{First moment}

We start with the first moment ---substituting $k=1$ in  Eq.~\eqref{eq:mks} we get,
\begin{align}
m_1(s) = \frac{a_0}2 \sum_{\sigma_1=\pm 1} \mu(\sigma_1) \sum_{n=0}^{\infty} \int_0^\infty \prod_{l=1}^{n} d\tau_l d \tau_{n+1} \rho(\tau_l) f(\tau_{n+1}) \exp{(-s\sum_{l=1}^{n+1} \tau_l)} \sum_{i,j=1}^{n+1}\chi_{ij} \tau_i \tau_j \label{eq:m1s_def}
\end{align}
where the symmetric matrix $\chi_{ij}$ is defined in Eq.~\eref{eq:defn_xchi}. To evaluate the above expression it is convenient to simplify the summand first. To this end, we separate the $i=j$ and $i\ne j$ and denote
\begin{align}
\frac 12 \sum_{i,j=1}^{n+1}\chi_{ij} \tau_i \tau_j  = S_1(n) + \frac 12 S_2(n),
\end{align}
where, we have defined,
\begin{align}
S_1(n) &= \sum_{i=2}^{n+1} \sum_{j=1}^{i-1} \sigma_j \tau_j \tau_i, \quad \text{and}, \quad
S_2(n) = \sum_{i=1}^{n+1} \sigma_i \tau_i^2 
\end{align}
Then, \eref{eq:m1s_def} can be expressed as,
\bea 
m_1(s) = a_0 \sum_{\sigma_1 = \pm 1} \mu(\sigma_1)\sum_{n=0}^\infty \left [ I_1(n) + \frac 12 I_2(n) \right], \label{eq:I1I2}
\eea
where,
\begin{align}
I_\nu(n) &=  \int_0^\infty \prod_{l=1}^{n}\id \tau_l \id \tau_{n+1} \rho(\tau_l) f(\tau_{n+1}) \exp{(-s\sum_{l=1}^{n+1} \tau_l)} S_\nu(n). \label{eq:Inu_m1}
\end{align}
To evaluate the integrals over $\{ \tau_i \}$ it is convenient first to simplify the sums $S_\nu (n)$, by separating the terms involving $\tau_{n+1}$ since it involves different weight. We then have,
\bea
S_1(n) = \sum_{i=2}^{n}\sum_{j=1}^{i-1} \sigma_j \tau_j \tau_i + \sum_{j=1}^n \sigma_j \tau_j \tau_{n+1}. 
\eea
Substituting the above equation in \eref{eq:Inu_m1} and performing the integrals over $\{\tau_i\}$ term by term, we have,
\bea 
I_1(n) &=& \sum_{i=2}^{n} \sum_{j=1}^{i-1} \sigma_j h(s) g'(s)^2 g(s)^{n-2}  + \sum_{j=1}^n \sigma_j h'(s) g'(s) g(s)^{n-1}, \label{eq:I1}
\eea 
where $g(s)$ is the Laplace transform of $\rho(\tau)$ [see Eq.~\eref{eq:g-def}], and $h(s)$ is the Laplace transform of the residual time distribution $f(\tau)$, given by,
\bea
h(s) = \int_0^\infty d \tau \, e^{-s \tau} f(\tau) = \frac 1 s (1 - g(s)). \label{eq:h-def}
\eea
Remembering that $\sigma_j = \sigma_1 (-1)^{j-1},$ the sum over $i,j$ in \eqref{eq:I1} can be immediately evaluated to yield,
\bea 
\fl I_1(n) &=& \frac {\sigma_1}4 g'(s) g(s)^{n-2} \Bigg[ \Big(2n-1+ (-1)^n \Big)g'(s) h(s)+ 2(1-(-1)^n) g(s) h'(s) \Bigg ].
\eea 
We proceed in a similar manner to evaluate $I_2(n)$. Separating the $i=n+1$ term in $S_2(n)$ and performing the $\{\tau_i \}$ integrals, we have,
\bea 
I_2(n) &=& \sigma_1 g(s)^{n-1} \Bigg[ \frac 12 (1- (-1)^n)g''(s) h(s) + (-1)^n g(s) h''(s) \Bigg ]
\eea 
Finally, performing the sum over $n$ in Eq.~\eqref{eq:I1I2} and using Eq.~\eqref{eq:h-def} we get the Laplace transform of the average position which is quoted in \eqref{eq:xav_laplace}.

\subsection{Second moment}

To compute the Laplace transform of the second moment, we put $k=2$ in Eq.~\eref{eq:mks},
\begin{align}
m_2(s) = \frac{a_0^2}{4} \sum_{\sigma_1 = \pm 1} \mu(\sigma_1)\sum_{n=0}^\infty \int_0^\infty \prod_{i=1}^{n}\id \tau_i \id \tau_{n+1} \rho(\tau_i) f(\tau_{n+1}) e^{-s\sum_{i=1}^{n+1} \tau_i} \left(\sum_{i,j=1}^{n+1} \chi_{ij}\tau_i \tau_j \right)^2 \label{eq:ap_m2s}
\end{align}
Before evaluating the integrals, it is convenient to simplify the summand in the integrand. To this end, let us denote,
\bea
q(n) \equiv  \frac 14 \left (\sum_{ij} \chi_{ij} \tau_i \tau_j \right )^2 = S_1(n) + S_2(n) + S_3(n),
\eea
where, we have defined,
\bea
S_1(n) &=& \frac 14 \sum_{i=1}^{n+1}\sum_{j=1}^{n+1} \sigma_i \sigma_j \tau_i^2 \tau_j^2, \cr
S_2 (n) &=&  \sum_{i=2}^{n+1} \sum_{j=1}^{i-1}\sum_{l=2}^{n+1}\sum_{m=1}^{l-1} \sigma_j \sigma_m \tau_i \tau_j \tau_l \tau_m, \cr
S_3(n) &=& \sum_{i=2}^{n+1} \sum_{j=1}^{i-1}\sum_{l=1}^{n+1} \sigma_j \sigma_l \tau_i \tau_j \tau_l^2.
\eea
From \eref{eq:ap_m2s}, the Laplace transform of the second moment can then be expressed as,
\begin{align}
m_2(s)=a_0^2 \sum_{\sigma_1 = \pm 1} \mu(\sigma_1)\sum_{n=0}^\infty \left(I_1(n) +I_2(n)+I_3(n)  \right),
\label{eq:m2sf}
\end{align}
where,
\begin{align}
I_\nu(n)=\int_0^\infty \prod_{i=1}^{n}\id \tau_i \id \tau_{n+1} \rho(\tau_i) f(\tau_{n+1}) e^{-s\sum_{i=1}^{n+1} \tau_i} S_\nu(n)
\label{eq:I_nu}
\end{align}

Next we simplify each of the sums further by exploiting symmetries. Furthermore, it is convenient to separate the terms involving $\tau_{n+1}$ since it comes with different trajectory weights.  

Let us first consider $S_1(n)$. Since the summand is symmetric under exchange of $i$ and $j$, we get, by separating the $i=j$ term,
\bea
S_1(n) = \frac 14 \Bigg[2 \sum_{i=2}^{n+1}\sum_{j=1}^{i-1} \sigma_i \sigma_j \tau_i^2 \tau_j^2 + \sum_{j=1}^{n+1}  \tau_j^4 \Bigg] \label{eq:S1}
\eea
where we have used the fact $\sigma_j^2=1.$ Next we need to calculate $I_1(n)$ [see \eref{eq:I_nu}], where, to perform the integrals over $\tau_i$, it is convenient to separate the $i=n+1$ terms in \eref{eq:S1}. Then, one can perform the integral term by term to get,
\begin{align}
I_1(n)=& \frac 12 \Big[\sum_{i=2}^{n}\sum_{j=1}^{i-1} \sigma_i \sigma_j g''(s)^2 g(s)^{n-2} h(s) +   \sum_{j=1}^n \sigma_{n+1} \sigma _j g''(s) g(s)^{n-1} h''(s) \Big]\cr
 & + \frac 14 \Big[n g''''(s) g(s)^{n-1} h(s)+ g(s)^n h''''(s) \Big], 
\end{align}
where, $g(s)$ and $h(s)$ are  the Laplace transforms of the waiting time and residual time distrinutions $\rho(\tau)$ and $f(\tau)$, respectively [see \eref{eq:g-def} and \eqref{eq:h-def}]. Remembering that $\sigma_i = (-1)^{i-1} \sigma_1$ the sums over $i$ and $j$ can also performed directly, leading to,
\begin{align}
I_1(n)= & \frac 18  (1-2n-(-1)^n) g''(s)^2 g(s)^{n-2} h(s) +\frac 14 \Big[((-1)^n-1) g''(s) g(s)^{n-1} h''(s) \cr
& + n g''''(s) g(s)^{n-1} h(s)+ g(s)^n h''''(s) \Big].  
\end{align}
Note that, unlike the first moment, here there is no dependence on $\sigma_1$. 

Next, we consider the second sum $S_2(n)$. In this case, the summand is symmetric under the exchange of the indices $i$ and $l$. Hence it is convenient to  separate the terms with $i=l$, and write
\bea
S_2(n) = 2 \sum_{i=3}^{n+1} \sum_{l=2}^{i-1}\sum_{j=1}^{i-1}\sum_{m=1}^{l-1} \sigma_j \sigma_m \tau_i \tau_j \tau_l \tau_m + \sum_{i=2}^{n+1} \sum_{j=1}^{i-1}\sum_{m=1}^{i-1} \sigma_j \sigma_m \tau_i^2 \tau_j \tau_m
\eea
We proceed further by exploiting the symmetries. For the first term, we first separate the sum over $j$ in three components, namely, $1 \le j \le l-1$, $j=l$ and $ l+1 \le j \le i-1$. For the second term, which is symmetric in $j,m$
we again separate $j=m$ term and proceed as before. Finally, we arrive at an expression where all the terms involve only ordered sums,
%with the same value of two or more indices have been separated, which is given by,
\begin{align}
S_2(n) = & 2 \sum_{i=4}^{n+1} \sum_{l=3}^{i-1} \sum_{j=2}^{l-1}\sum_{m=1}^{j-1} (2 \sigma_j \sigma_m + \sigma_l \sigma_m) \tau_i \tau_j \tau_l \tau_m  \cr
 + & 2 \sum_{i=3}^{n+1} \sum_{l=2}^{i-1}\sum_{j=1}^{l-1} (\tau_i \tau_j^2 \tau_l + \sigma_l \sigma_j (\tau_i \tau_j \tau_l^2 + \tau_i^2 \tau_j \tau_l)) 
+ \sum_{i=2}^{n+1}\sum_{j=1}^{i-1} \tau_i^2 \tau_j^2.
\end{align}
Proceeding in the same manner as before, we also compute $I_2(n)$ [see \eref{eq:I_nu}], which leads to a rather long expression,
\begin{align}
I_2(n) =& \frac 1{24} \Big[15 - 58 n + 42 n^2 - 8 n^3 + 3(-1)^n(2n -5)\Big] g'(s)^4 g(s)^{n-4} h(s) \cr
+ &\frac 12 (n-2)\big(1-2n- (-1)^n \big) g'(s)^3 g(s)^{n-3} h'(s) \cr 
+ & \frac 16 \Big[-3 + 2n(n-2)(n-4)+ 3 (-1)^n \Big] g'(s)^2 g''(s) g(s)^{n-3} h(s) \cr
 +& \frac 12 (1-2n-(-1)n) g'(s)^2 g(s)^{n-2}h''(s) \cr
+&  \frac 12 n(n-1) g''(s)^2 g(s)^{n-2} h(s) + n g''(s) g(s)^{n-1} h''(s)
\end{align}

Finally, for the last term $S_3(n)$, we separate the sum over $l$ into three parts, namely $l<i$, $l=i$ and $l>i$,
\bea
\fl S_3(n) = \sum_{i=2}^{n+1}\sum_{j=1}^{i-1}\sum_{l=1}^{i-1} \sigma_j \sigma_l \tau_i \tau_j \tau_l^2 + \sum_{i=2}^{n+1}\sum_{j=1}^{i-1} \sigma_i \sigma_j \tau_i^3 \tau_j + \sum_{i=2}^{n+1}\sum_{j=1}^{i-1}\sum_{l=i+1}^{n+1} \sigma_j \sigma_l \tau_i \tau_j \tau_l^2 
\eea
We proceed in the same method outlined above, exploiting symmetries of the terms with the objective of arriving at terms involving only ordered sums. After a few steps of algebra we get,
\begin{align}
S_3(n) = \sum_{i=3}^{n+1}\sum_{j=2}^{i-1}\sum_{l=1}^{j-1} (2 \sigma_j \sigma_l \tau_i \tau_j \tau_l^2 + \sigma_i \sigma_l \tau_i^2 \tau_j \tau_l) + \sum_{i=2}^{n+1}\sum_{j=1}^{i-1} (\tau_i \tau_j^3 + \sigma_i \sigma_j \tau_i^3 \tau_j).
\end{align}
Next, we compute $I_3(n)$, which turns out to be,
\begin{align}
I_3(n) =& \frac 14 (n-2)(1-2n-(-1)^n) g'(s)^2 g''(s)g(s)^{n-3}h(s) \cr
+& \frac 12 (1- 2 n -(-1)^n) g'(s)g''(s)g(s)^{n-2} h'(s) + \frac 14 \big(1 -(-1)^n (1-2n)\big) g'(s)^2 g(s)^{n-2} h''(s)\cr
+& \frac 14 (7-8n+2 n^2 +(-1)^n) g'(s) g'''(s) g(s)^{n-2}h(s) + n g'''(s) g(s)^{n-1}h'(s) \cr
-& \frac 12 (1-(-1)^n) g'(s) g(s)^{n-1}h'''(s).
\end{align}

Finally, combining the three different contributions $I_{1,2,3}(n)$, performing the sum over $n$ in \eref{eq:m2sf}, and using \eref{eq:h-def},  we arrive at a simple expression for $m_2(s)$, which is quoted in \eref{eq:x2av_laplace}.

\ack
I.S. acknowledges the Student Associateship program at S. N. Bose Centre for Basic Sciences, India. D.A. acknowledges support from the Department of Science and Technology, India through a Kishore Vaigyanik Protsahan Yojana (KVPY) fellowship. U.B. acknowledges support from Science and Engineering
Research Board (SERB), India, under a Ramanujan Fellowship (Grant No. SB/S2/RJN-077/2018).

\end{document}